\documentclass[aps,prl,preprint,groupedaddress]{revtex4}
\usepackage{amssymb,amsmath,amsbsy}
\usepackage{mathrsfs}
\usepackage[dvips]{graphics}
\usepackage{epsfig}

\def\be{\begin{eqnarray}}
\def\ee{\end{eqnarray}}
\def\bea{\begin{eqnarray*}}
\def\eea{\end{eqnarray*}}

\def\centeron#1#2{{\setbox0=\hbox{#1}\setbox1=\hbox{#2}\ifdim
\wd1>\wd0\kern.5\wd1\kern-.5\wd0\fi
\copy0\kern-.5\wd0\kern-.5\wd1\copy1\ifdim\wd0>\wd1
\kern.5\wd0\kern-.5\wd1\fi}}
\def\ltap{\;\centeron{\raise.35ex\hbox{$<$}}{\lower.65ex\hbox{$\sim$}}\;}
\def\gtap{\;\centeron{\raise.35ex\hbox{$>$}}{\lower.65ex\hbox{$\sim$}}\;}

\newcommand{\newc}{\newcommand}
\newc{\qbar}{{\overline q}}
\newc{\Kahler}{K\"ahler }
\newc{\deltaGS}{\delta_{\rm GS}}

\begin{document}
\preprint{
\vbox{\vspace*{2cm}
      \hbox{UCI-TR-2010-12}
      \hbox{June, 2010}
}}
\vspace*{3cm}

\title{Dirac Gauginos, Negative Supertraces and Gauge Mediation}
\author{Linda M. Carpenter}

\affiliation{Department of Physics and Astronomy  \\
   University of California Irvine, Irvine, CA U.S.A. \\
   lcarpent@uci.edu \\
\vspace{1cm}}

\begin{abstract}
In an attempt to maximize General Gauge Mediated parameter space, I propose simple models in which gauginos and scalars are generated from disconnected mechanisms.  In my models Dirac gauginos are generated through the supersoft mechanism, while independent R-symmetric scalar masses are generated through operators involving non-zero messenger supertrace.  I propose several new methods for generating negative messenger supertraces which result in viable positive mass squareds for MSSM scalars.  The resultant spectra are novel, compressed and may contain light fermionic SM adjoint fields.

\end{abstract}


\maketitle

\section{Introduction}

Gauge mediation is a simple and flavor blind mechanism for the communication of SUSY breaking \cite{Dine:1993yw} \cite{Dine:1994vc}
\cite{Dine:1995ag}.  In gauge mediation, SUSY breaking is communicated from a hidden sector to the MSSM via SM gauge interactions only.  In its simplest implementation,
gauge mediation requires a SUSY breaking spurion which gets an F term as well as an R symmetry-breaking vev.

This spurion couples to a set of fields with SM gauge interactions, the
messengers generating a superpotential
\be
W = X M \overline{M}\rightarrow v M\overline{M} + \theta^2 F_X M\overline{M}
\ee
X is the SUSY breaking spurion and M are the messengers, here in a fundamental anti-fundemental of the SM gauge group SU(5).  Loops proportional to SM gauge couplings are then generated giving masses to the MSSM fields.  Majorana gauginos masses are generated at one loop while scalar masses squared are generated at two loops.   In the minimal case, all MSSM fields get masses proportional to a single mass parameter $\Lambda = F/v$ .  The theory is thus simple and predictive, however, like many minimal SUSY breaking models,
Minimal Gauge Mediation(MGM) is highly constrained as it produces a hierarchichal spectrum with very heavy squarks. There is thus significant fine tuning in the Higgs potential.  Meade \emph{et. al.} have proposed a generalization of gauge mediation, defining Gauge Mediation as any mediation mechanism such that MSSM masses go to zero when the SM gauge couplings are turned off.  Barring A and B terms, a model within the framework of General Gauge Mediation(GGM) may have up to six distinct mass parameters\cite{Meade:2008wd}.  This framework allows gauge mediated models with non-standard spectra and the hope of less fine tuning.   Simple GGM extensions have been built, however not every model manages to cover the entire GGM parameter space.

There are many regions of MSSM parameter space which produce highly non-standard spectra with rich phenomenology.  Many of these regions of parameter space are not yet highly constrained by LHC searches, for example 'supersoft' spectra \cite{Kribs:2012gx}, compressed spectra, and -strikingly - spectra with stop masses under 400 GeV \cite{stops}- \cite{stops5}. The General Gauge Mediated framework should allow near complete models which express these spectra to be built.
 
However, many simple weakly coupled models retain spectra very similar to Minimal Gauge Mediation. In particular, squarks
remain the heaviest sparticles with masses 500 GeV or above due to a persistent relation between gaugino and scalar masses.
To get the gist of this consider an example from a set of weakly coupled renormalizable models \cite{Carpenter:2008wi}.  For a model containing messenger pairs in a
$5$, $\overline{5}$ and $10$, $\overline{10}$ of $SU(5)$ and for multiple SUSY breaking spurions, the MSSM mass spectrum is given by five parameters. However, the large number of parameters
does not necessarily guarantee light scalars.  In particular gluino masses are given by the sum of three mass parameters
\be
m_{g} = \frac{\alpha_3}{4\pi}(\Lambda_q + 2\Lambda_Q +\Lambda_u)
\ee

One is free to cancel the mass parameters to produce an arbitrarily light gluino.  The largest contribution to the squark mass, however, is given
by a mass parameter which is the \emph{sum} of the squares of the gluino mass parameters
\be
m_{s}^2 \sim \frac{\alpha_3}{4\pi}^2 \Lambda_c^2 \rightarrow \frac{\alpha_3}{4\pi}^2(\Lambda_q^2 + 2\Lambda_Q^2 +\Lambda_u^2)
\ee

The squarks may not be made arbitrarily light with respect to the gluinos, in fact many small gluino masses which rely on a cancelation between
large mass parameter will ensure a heavy squark mass.  Given the details of tuning in the EWSB sector, this feature of weakly
coupled models leads to the general occurrence of an irreducible fine tuning at the 5 percent level \cite{Carpenter:2008he}.
As Seiberg \emph{et. al.} have shown, breaking these mass relations requires more complex messenger sectors \cite{Buican:2008ws}.
In many models with hidden sector gauge dynamics mass relations between scalars and gauginos still persist; for example, in current semi-direct gauge
mediated models there is an irreducible bound on the ratio of scalars and gaugino masses \cite{Dumitrescu:2010ha}.

In attempting to cover gauge mediated parameter space, there are unexplored theoretical options.  In this paper I will generate scalar and
gaugino masses from disconnected mechanisms.  My models require messenger fields with non-holomorphic masses and a hidden sector gauged
$U(1)$ field which gets a D-term vev. Integrating out sets of messengers will produce two distinct kinds of MSSM masses, Dirac gaugino masses generated using
the supersoft mechanism, and  R-symmetric, log-divergent scalar masses.  In particular I propose several new mechanisms to generate positive
scalar mass squareds from negative messenger supertraces.

Dirac gauginos have  proven theoretically useful, yielding mediation mechanisms such as supersoft SUSY
 breaking\cite{Fox:2002bu}, supersoft hybrids \cite{Carpenter:2005tz}, gaugino mediation \cite{Kaplan:1999ac} and the MRSSM \cite{Amigo:2008rc}.
 Models in this paper will employ the supersoft mechanism.
Supersoft SUSY breaking yields a Dirac gaugino mass when gauginos mix with an additional SM adjoint field \cite{Fox:2002bu}; masses are proportional to a
SUSY breaking D-term \cite{Fox:2002bu}.  Generating Dirac gauginos through gauge mediation was proposed by Nelson\emph{ et. al.},  \cite{Fox:2002bu}\cite{Nelson:2002ca},
and larger explorations of  Dirac gauginos in the formalism of general gauge mediation have been made \cite{Benakli:2008pg}.
Supersoft type gauginos offer several model building advantages.  First, they  naturally allow three distinct gaugino mass parameters with minimal
theoretical structure. Second, the scalar mass contribution resulting from supersoft generates scalars masses a square root of a loop factor
below gaugino masses.

When supersoft is the only mass giving mechanism, producing a 100 GeV mass spectrum for MSSM scalars requires multi-TeV gaugino masses.
Since the aim in this paper is to disconnect gaugino and scalar masses however, one is free to generate 100 GeV gaugino masses from supersoft mediation.  The bulk of the scalar masses are generated not from supersoft, but from the
independent R symmetric contributions which do not affect gaugino masses.  In this way the gaugino and scalar masses become largely independent of one another.   In addition I resolve a problem found
in previous GM implementations of supersoft in which large negative mass squared contributions to adjoint scalar masses were generates.  In my
models, adjoint scalar mass squared are large and positive.  The existence of SM adjoints
offers novel solutions to old theoretical problems such as the $\mu$ term-less MSSM \cite{Nelson:2002ca}.

The effect of non-holomorphic messenger masses on the MSSM scalar spectrum was first discussed by Poppitz and Trivedi \cite{Poppitz:1996xw}.
 When messengers have non-holomorphic masses they result in  two loop gauge mediated
 masses for MSSM scalars, that are proportional to the messenger supertrace.  As these scalar mass operators are R symmetric, they do not affect gaugino masses.  The catch to using this mechanism
 to generate scalar masses is that
the scalar mass squareds have the opposite sign as the messenger supertrace.   Many models are known to generate \emph{positive} messenger supertrace and hence
disastrous \emph{negative} messenger mass squareds, however few methods are known to generate negative messenger supertrace. One of the few methods for generating
negative messenger supertrace was proposed by Randall \cite{Randall:1996zi}.  This
'mediator mechanism' involved generating two-loop non-holomorphic messenger masses.  I build several models where negative messenger supertrace
is achieved at lower loop order.

This paper proceeds as follows, section 2 reviews supersoft mediation.  Section 3 presents a SUSY breaking superpotential that generates Dirac
gauginos.  Section 4 adds to this superpotential terms which are capable of generating R symmetric MSSM scalar masses through
negative  messenger supertrace.  Section 5 concludes.

 \section{Review of Dirac Gauginos}

In Supersoft SUSY breaking scenarios, gaugino masses are Dirac.  In the simplest case, they arise from the coupling of SM gauge adjoints
to gauginos and a hidden sector $U(1)^{'}$ gauge field which gets a D term.  One must add to the MSSM three adjoint fields, one for each gauge group.  In the low energy, the super-potential operator generating such mass
terms is
 \be
 W = c_i\frac{W^{'}W_i A^i}{\Lambda}
 \ee
 where the A is the adjoint and $W^{'}$ is the hidden sector U(1) gauge field.  Gauge indicies are contracted
 between $W$ and $A$ while Lorentz indicies are contracted between $W$ and $W^{'}$. Inserting the D term and pulling the proper components the above
 expression  becomes

 \be
c_i \frac{D}{\Lambda}\lambda_i \psi_{Ai}
 \ee
which is a  Dirac gaugino mass for gauginos of size $c_i{D}/{\Lambda}$.  As the couplings $c_i$ are different for all of the gauginos,
each gaugino is determined by a distinct mass parameter.

Finite one loop scalar masses are then generated after gauginos get mass.  In the absence of extra operators, the scalar masses are
\be
m_s^2= \frac{C_i \alpha_i {m_{\lambda i}}^2}{\pi} \log (\frac{\delta_i }{ m_{\lambda i}})^2
\ee
where $m_{\lambda i}$ are the gaugino masses and $\delta_i$ is the mass squared of the real part of the adjoint.  The ratio between the masses of the
gauginos and the MSSM scalars (with the exception of the adjoints) is thus
 \be
 \frac{m_s}{m_\lambda}= \sqrt{\frac{2 C_i\alpha_i }{\pi}\log (\frac{\delta_i }{ m_{\lambda i}})}
 \ee
Which means that scalar masses from the supersoft  mechanism may be a few percent of the gaugino masses.  If gauginos masses are fixed at
100 GeV and there are R symmetric contributions to scalar masses, then the gaugino and scalar masses are largely independent.
It is interesting to note that if $\delta_i$ equals $m_{\lambda i}$, the gaugino masses make no contribution to scalar masses at all.

 \section {SUSY breaking and Dirac Gauginos}
 I now attempt to implement supersoft mediation by building  a hidden sector with a $U(1)^{'}$ D-term that couples to an adjoint and gaugino.
 This is accomplished by integrating out messengers which are charged both under the hidden sector U(1) and the standard model gauge interactions.
Consider the following hidden sector superpotential with a gauged U(1) symmetry

\be
W= \lambda X (\phi_{+}\phi_{-}- \mu^2) + m_1\phi_{+}Z_{-} + m_2 \phi_{-}Z_{+}+ W^{'}W^{'}
\ee

The subscripts indicate $U(1)^{'}$ charges.  This model was studied in \cite{Dine:2006xt}.  The fields $\phi$ get $U(1)^{'}$
breaking vevs

\be
\phi_{+}^2=\frac{m_2}{m_1}\phi_{-}^2
\ee
\be
\phi_{-}= \sqrt{\frac{m_1}{m_2}\mu^2-m_1^2} \nonumber
\ee
The field X as well as the Z's get F terms of order $m/\lambda$.

The D term is nonzero as long and $m_1$ is unequal to $m_2$, and is given by
\be
D = g^{'}(\frac{m_1}{m_2}\mu^2-m_1^2)(\frac{m_2}{m_1}-1)
\ee
The coupling of the D term to the gaugino and adjoint requires the addition of a set of messengers in the fundamental representation of the SM
gauge groups, and which
are also charged under the $U(1)^{'}$.  The messengers require a supersymmetric mass-term.
Adopting the notation of Nelson, the messenger superpotential is given by

\be
W_T= m_T T\overline{T}+ y_i\overline{T}AT
\ee
where the fields T are the messengers.  At tree level there are no off-diagonal bosonic messenger masses but there are diagonal scalar messenger masses
resulting from the D term.  Since the messenger fundamental and anti-fundamental have opposite
$U(1)^{'}$ charge for anomaly cancelation, the supertrace of these messengers is zero.  A one loop mass for gauginos is generated and, is given by the diagram found below.  The resulting gaugino mass is

\be
m_{\lambda_i} = \frac{g_i}{16\pi^2}\frac{y_i D}{m_T}
\ee

Some notes are in order.  First, one may call this implementation of supersoft a  sub-set of semi-direct gauge mediation; that is the messengers
have charge under a hidden sector
gauge group but do not themselves participate in SUSY breaking \cite{Seiberg:2008qj}. In this model there is a flat direction, and one may do a
Coleman-Weinberg calculation to lift it.  For certain values of the parameters R symmetry is broken, and $X$ gets a non zero vev.  One may find a
vev for $X$ in the region of large $m_1$ and $m_2$ for order 1 values of $\lambda$ and the gauge coupling g.  Here an $X$ vev means that a messenger B-term is generated at two loops and hence a Majorana gaugino mass is generated at three loops.

The one loop R-symmetric Dirac mass will by far be the dominant gaugino mass contribution and the three loop mass will be small.
A multi-loop gaugino mass seems generic for models where messengers are charged under the hidden sector gauge group but do not participate in
SUSY breaking; both the Mediator Models of Randall and Semi-Direct Gauge Mediation have this feature.

\subsection{Operators from the Kahler Potential}

In addition to gaugino masses, important mass contributions to the real and imaginary part of the adjoint are generated at one loop.
As the original superpotential contains neither the supersoft gaugino mass nor the adjoint mass contributions, one must extract them from the
Kahler potential.  In order to keep careful track of one loop operators, I will now produce them from the Kahler potential, including a
rederivation of the supersoft gaugino mass.

\begin{figure}[h]
\centerline{\includegraphics[width=7 cm]{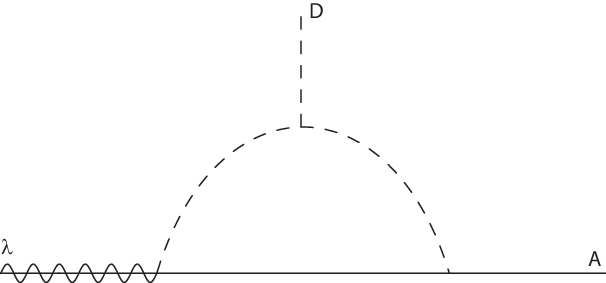}}
\caption{\label{diag1}1-loop diagram for gaugino masses.}
\label{fig:2pgm}
\end{figure}

  The Kahler potential term which generates the one loop supersoft gaugino mass is
\be
K=\int d^4 \theta \frac{WDV^{'}A}{\Lambda} + h.c.
\ee
Where V is the $U(1)^{'}$ vector field, W is the SM field strength and A is the adjoint.  In this operator gauge indicies are contacted between W and
A, while Lorentz indicies are contracted between W and the superspace derivative D.  Acting with the superspace derivative, integrating over
$\overline{\theta}$, and replacing $V{'}$ by its vev one gets
\be
W=\int d^2 \theta \frac{D^{'}}{\Lambda}WA
\ee
which is exactly the supersoft term in the superpotential.

There are other operators of the same order only involving the adjoint field.  These operators are
\be
K= \int d^4 \theta \frac{W^{'}DV^{'}AA}{\Lambda^2}+ h.c. \nonumber
\ee
and
\be
K= \int d^4 \theta \frac{W^{'}DV^{'}AA^{\dagger}}{\Lambda^2}+ h.c.
\ee
which are soft mass contributions to the real and imaginary parts of the adjoint A.  These terms are represented by two one loop diagrams
involving messengers which are given in the Appendix.  If one takes the first operator and acts with derivatives, integrates over $\overline{\theta}$,
 and inserts D-terms, the result is a term in the superpotential
 \be
W=\int d^2 \theta \frac{D^{'}D^{'}}{\Lambda^2}AA
\ee
This operator was discussed by Nelson and is recognizable as the 'lemon twist' operator; in fact presented a challenge for previous incarnations of Supersoft mediation.  This and similar operators can be quite useful for model building, for example \cite{Carpenter:2005tz} \cite{Carpenter:2005jj}.  In this context, however, the operator is problematic because it gives a negative mass
squared contribution to one component of A.  This mass contribution is large, in fact it is a square root of a loop factor larger than the
gaugino mass.  If only this first operator existed, one would require a very large majorana mass for A in order to
preserve the SM gauge symmetries.  However, the second operator is of the same order as the first and gives both components of A positive mass squared.
Adding up the contributions to A one indeed finds a cancelation, to order $D^2/M^2$ to real part of A  while the imaginary
gets a positive mass squared.  The full scalar adjoint masses consists of the D term-mass for the real component,  the above one loop
masses, and whatever R symmetric gauge mediated masses the adjoints may get.  One expects the gauge mediated masses will be large and positive and
both components of A thus end up with large positive mass squareds.  Operators of the type found in equation 15 seem to be generic to GM models of Dirac gauginos, for example one may notice that slightly reminiscent operators may be found in \cite{Amigo:2008rc}.
In calculating masses, one will only find all relevant operators by keeping careful track of the Kahler potential, keeping track of the superpotential alone is not sufficient.

\section{Scalar Masses and Negative Messenger Supertraces}

One may create R symmetric scalar masses which are independent of the gaugino masses.  The scalar mass contributions arise at two loops from diagrams where messengers have non-holomorphic masses.  Such terms are log divergent and proportional to
the messenger supertrace.  The MSSM
scalar masses from such contributions were calculated by Poppitz and Trivedi \cite{Poppitz:1996xw} and are given by

\be
m^2_i = - f\sum_a\frac{g_a^4}{128 \pi^4} S_Q C_{ai} {\rm Str} M_{mess}^2 \log(\frac{M^2}{\Lambda^2})
\ee
where S is the Dynkin index of the messengers, $C_{ai}$ is the Casimir for the scalars, and M is the supersymmetric messenger mass.  These mass
terms come from R symmetric physics and hence do not effect gaugino masses.  The scalar mass squareds have signs which are opposite of the
sign of messenger supertrace; \emph{positive} scalar mass squareds require \emph{negative} messenger supertraces.

One of the only fleshed out examples of negative messenger supertrace are the Mediator Models of Randall \cite{Randall:1996zi}.  In this model, there are low scale messengers which are charged under SM gauge groups, and high scale messengers-uncharged under SM gauge groups- which talk to the SUSY breaking sector.
The argument is as follows, the hidden sector messengers get SUSY breaking masses though direct mediation, and thus have positive supertrace.
The low scale messengers, in addition to having a supersymmetric mass, get two-loop SUSY breaking masses from the high scale messenger sector.  Since the high scale messengers had positive
supertrace, they will contribute negatively to the SUSY breaking masses of the low scale messengers. Thus the low scale messengers get a negative
supertrace which results in a positive mass squared for MSSM scalars.  The entire mechanism generates MSSM scalar masses at four loops, but one would like to achieve scalar masses at lower loop level.

Achieving a negative messenger supertrace is challenging.  To get a  negative supertrace, the bosonic messenger mass squared must be less than the square
of the fermionic messenger mass.  Any supersymmetric mass does not contribute to the supertrace.  In addition, at tree level any F term type SUSY breaking leads to zero supertrace.

  One sees that the only way to
achieve negative supertraces at tree level is with D-terms. Messengers that couple to a D-term may get tree level gauge mediated mass squared which are
negative \cite{Nardecchia:2009ew}.

Getting negative messenger supertraces with F-terms alone requires that the messengers get nonsupersymmetric mass contributions at loop level.
   I will build several models with negative messenger supertrace at
different loop order levels, one with and one without hidden sector gauge dynamics.

 \subsection{Tree Level SuperTrace}
 To generate tree-level non-supersymmetric messenger masses I require a hidden sector gauge group which gets a D term.
For the sake of simplicity I will be using a hidden sector $U(1)^{'}$ gauge group. I will consider the superpotential from section 3
 \be
W= \lambda X (\phi_{+}\phi_{-}- \mu^2) + m_1\phi_{+}Z_{-} + m_2 \phi{-}Z_{+}
\ee
as it breaks SUSY and has a $U(1)^{'}$ D term.
I must now add the messenger content in such a way that the messengers pick up a non-zero supertrace from the D term.  This means that the
messengers will be spectators which are charged under the hidden sector $U(1)^{'}$ but will not participate in SUSY breaking.  Building a viable model with the
correct messenger sector has several constraints; first, the messenger content must be such that there are no $U(1)^{'}$ or mixed anomalies.  Second,
the messenger supertrace must actually be non-zero; if, as in the previous section,
one adds a pair of messengers $M$, $\overline{M}$ with opposite $U(1)^{'}$ charges and identical supersymmetric mass, the tree
level supertrace contributions to $M$ and $\overline{M}$ will exactly cancel.
  Finally, if messenger bi-linears have a $U(1)^{'}$ charge that does not sum to zero, they cannot be given an
 explicit mass term, their mass must come from the vev of a charged field.

 To meet the above requirements, consider a superpotential which is a variant of the one in section 3, with two sets of messengers M and N

\be
W= \lambda X (\phi_{+}\phi_{-}- \mu^2) + m_1\phi_{+}Z_{-} + m_2 \phi{-}Z_{+} + \lambda_2\phi_{+}M\overline{N} + \lambda_1\phi_{-}N\overline{M}
\ee

The fields M and $\overline{N}$ have charges which sum to 1 while  N and $\overline{M}$ have charges that sum to -1. For simplicity we may take
N and $\overline{N}$ to have $U(1)^{'}$ charge zero so $M$ and $\overline{M}$ have opposite charge.   The M's do not have equal supersymmetric mass.
 M has mass $\lambda_2v_{\phi_{+}}$ and will make a contribution to the supertrace proportional to +D,  while $\overline{M}$  has a mass
  $\lambda_1v_{\phi_{-}}$ and will make a contribution to the supertrace proportional to -D.   However, the contributions to MSSM scalar masses
 do not exactly cancel since the supersymmetric messenger masses are different.  Applying the
Poppitz Trivedi formula for scalar masses one gets
\be
m^2_s = - f\sum_a\frac{g_a^4}{128 \pi^4} S_Q C_{ai} D \log(\frac{M_1}{\Lambda})+  f\sum_a\frac{g_a^4}{128 \pi^4} S_Q C_{ai} D \log(\frac{M_2}{\Lambda})
\ee
Summing the two non-canceling contributions yields

\be
m^2_i = - f\sum_a\frac{g_a^4}{128 \pi^4} S_Q C_{ai} D \log(\frac{M_1}{M_2})
\ee
where $M_1$ and $M_2$ are the unequal supersymmetric messenger masses.

As one expects, the scalar mass squareds go to 0 as supersymmetric messenger masses become degenerate. One may
understand the scalar mass formula as follows; both the messenger fundamental and anti-fundamental contribute to the scalar mass from running.  The
contributions to scalar masses from $M$ and $\overline{M}$ have opposite signs and would cancel if their supersymmetric mass thresholds were equal.  Since
the thresholds are unequal however, a scalar mass squared remains which is proportional to the mismatch of the running of both contributions.
  The sign of the mass squared depends on the relative size of the parameters $M_1$ and $M_2$; it can be made both negative and arbitrarily small.

To the above superpotential one may easily add the messenger sector of section 3.4 for a complete superpotential,
\be
\lambda X (\phi_{+}\phi_{-}- \mu^2) + m_1\phi_{+}Z_{-} + m_2 \phi{-}Z_{+} + \lambda_2\phi_{+}M\overline{N} + \lambda_1\phi_{-}N\overline{M} +
m_T T\overline{T}+ y_i\overline{T}AT
\ee
 Since the messengers  T have equal and opposite $U(1)^{'}$ charges and equal mass,
they will not effect the scalar mass relation just derived.  The messengers $M$ and $N$ do not couple to the adjoint and hence
do not contribute to the gaugino masses.  The ratio $\frac{M_1}{M_2}$ that appears in the scalar mass  is a parameter completely separate from
the gaugino sector.

In order to cover more scalar mass parameter space, one may divide the complete $SU(5)$ messenger multiplets
M into fundamentals of $SU(3$) and $SU(2)$ giving each separate couplings to the fields $\phi_+$ and $\phi_-$.

\be
W= \lambda_{Q2}\phi_{+}Q\overline{Y}+ \lambda_{Q1}\phi_{-}Y\overline{Q}+\lambda_{L2}\phi_{+}L\overline{E}+ \lambda_{L1}\phi_{-}E\overline{L}
\ee

There are now two mass parameters given by the ratio of the couplings $\lambda_{Q2}/\lambda_{Q1}$ and $\lambda_{L2}/\lambda_{L1}$.
Determining which coupling in the ratio is larger determines the sign of the log in the scalar mass formula.  One is thus free by choice of couplings
to give negative mass contribution to one set of scalar mass squared and positive contribution to another.  Therefore, even if one scalar, say the
squark, were to get large mass squared contribution from supersoft, it may be canceled by the log divergent mass while still maintaining positive mass
squared for the other scalars. Much of GGM parameter space is thus accessible.

\subsection{Loop Level SuperTrace}

Negative messenger supertraces may also be generated at loop level using only F term SUSY breaking. Using F terms I will attempt generate messenger
 masses at one loop, and thus MSSM scalar masses at three loops.

The F-term method of generating one loop negative mass squareds for messengers is reminiscent of that used to generate one loop MSSM scalar masses through
messenger-matter mixing \cite{Dine:1995ag}.  In the following model messengers start with supersymmetric masses.  They then mix with the
fields that talk to SUSY breaking.  As the result of mixing, the messengers get one loop scalar masses squared contributions which are negative,
 resulting in negative supertrace.

In standard messenger-matter mixing the SUSY breaking sector talks to intermediary fields, in this case the messengers, and the messengers would mix with Higgses which then get negative mass squareds.
In my scenario the role of the Higgses is played by the messengers and the role of the messengers is played by new intermediary fields.  I thus require a superpotential in which messengers only talk to SUSY breaking spurions through mixing with the new intermediary fields.

For two sets of messengers, M and N, in the fundamental and anti-fundamental reps. of the SM gauge groups, the required superpotential is

\be
W=X(H^2-\mu^2)+ m HA + y_1HM\overline{N}+ y_2H N\overline{M}
\ee

Where H and A are intermediary fields.  Here H gets a vev of order $\mu^2-m^2$ and X gets an F term.  The vev of H generates the supersymmetric messengers
masses.  The resulting  scalar potential
contains contributions
\be
V_s= y_1^2(h^2m^2+h^2\overline{n}^2)+y_2^2(h^2\overline{m}^2+h^2n^2)
\ee

This scalar potential leads to one loop diagrams for messenger masses which have H in the loop and insertions of the F term on the internal H lines.  In analogy to matter-messenger mixing, there are two diagrams that lead to mass squared contributions with opposite signs.

When calculating the resulting bosonic messenger masses there is an accidental cancelation in the highest order terms in F,  in this case
$\frac{F^2}{M^2}$.  This means the scalar messenger mass squareds
are given by the next highest order term proportional to $F^4$

\be
m_s^2 \sim -\frac{y^2}{16\pi^2}\frac{F^4}{M^6}
\ee
The negative mass squared contribution to the bosonic messenger ensures a negative messenger supertrace.
On expects that this messenger supertrace can not be
made positive and it will always result in positive mass squared contributions to the MSSM scalars given by

\be
m^2_i =  f\sum_a\frac{g_a^4}{128 \pi^4} S_Q C_{ai} \frac{y^2}{16\pi^2}\frac{F^4}{M^6} \log(\frac{v_H}{\Lambda})
\ee

Notice that the above superpotential has an R-symmetry.  Its structure is also non-generic and requires and additional $Z_2$ under which A, H,
and certain messengers have negative parity.  The superpotential only contains fields with R charge 0 or 2.  By a theorem \cite{Shih:2007av} we
expect that once
a Coleman-Weinberg calculation is made, the field X will get a vev at 0 and R symmetry will remain unbroken. If there were no
additional source of SUSY breaking this would mean a massless gaugino.  However, in the full model the gaugino mass  is covered since there is an independent SUSY
breaking source generating Dirac gaugino masses.

\section{Conclusions}

To cover General Gauge Mediated parameter space I have constructed models with Dirac gauginos and independent scalar masses from R-symmetric dynamics.
These models allow for a breaking of the relations between scalar and gaugino mass parameters and present the particular phenomenological possibility of
 light squark masses and a very degenerate sparticle spectrum.  In implementing Dirac gaugino masses I have avoided previous phenomenological problems with adjoint masses.
 In addition, I have proposed several new methods for generating negative messenger supertraces. Though I have presented simple viable models here there is still much work to do,
 in particular unification scenarios present a serious challenge.

While I have chosen very simple hidden sectors which contain the $U(1)^{'}$ field gauge  needed for Dirac gauginos, one
may pick hidden sector with more complicated dynamics, for example the 4-1 model.  In using D-terms to generate R-symmetric scalar masses I 
chose the D-terms from a $U(1)^{'}$ gauge symmetry, while it is likely possible to use D-terms from non-abelian gauge groups.   In addition it may be possible to marry gauge-messenger
 scenarios to the ideas I have presented here \cite{Intriligator:2010be}.  
 
One experimental challenge for Gauge Mediation of any kind is the possible discovery of a Higgs Boson at a mass 126 GeV.  In typical implementations of the MSSM it the Higgs mass is lifted above its tree level value through large one loop corrections from stop masses which must be above several TeV in mass which some may find phenomenologically unappealing.  Lighter stop masses may yield a Higgs mass of 126 GeV if large A terms are present(see for example \cite{Christensen:2012ei}) however minimal gauge mediation produces only small A terms which run under the gaugino mass, and models with dirac gauginos produce no A terms at all.

One should not despair however, the Higgs sector of MSSM models is incomplete until a mechanism can be specified for generating a $\mu$ term and a suitable $B_{\mu}$ term.  Any theoretical additions made to the Higgs sector may well raise the Higgs mass at tree level or loop level.  One example is the generation of large stop A-terms resulting from models with Higgs-messenger mixing \cite{Giudice:1997ni}, \cite{Craig:2012xp}.  Another example of course is the NMSSM.  It may also be possible to raise the Higgs mass significantly in models with Dirac gauginos, by coupling adjoint fields to the Higgs as is done in the $\mu$-less MSSM \cite{Carpenter}.  It is likely that many theoretical completions to General Gauge Mediated models exist which satisfy the (probable) Higgs mass bound.

In building these models I essentially required two disconnected SUSY breaking sectors which determine the masses of the scalars and gauginos respectively.  There may be a slight hierarchy between the SUSY breaking scales depending on the the choice of method of the generation of scalar masses, however the two SUSY breaking scales are generally comparable.  Though one may rely on anthropic arguments to explain a coincidence of SUSY breaking scales \cite{Cheung:2010mc} (footnote 14), it may not be beyond belief that such a coincidence could naturally take place.   Most models of General Gauge Mediation rely on the existence of multiple SUSY breaking spurions which have similar mass parameters, in addition to multiple sets of messengers - see for example \cite{Carpenter:2008wi}.  In principle these models are quite similar, it should not be too unnatural to choose similar mass parameters in the SUSY breaking sectors which lead to comparable masses for gauginos and scalars.  However, for those unsettled by the the reliance on the non-renormalization of the Superpotential, it may require some invocation of symmetries to explain why the two SUSY breaking sectors are sequestered from each other.

Finally, models with Dirac gauginos offer new fields with SM charges that make for interesting collider phenomenology. In standard
supersoft mediation, since gaugino masses come a loop factor above scalar masses, gauginos and adjoints are typically heavy, usually to 10 TeV to
produce scalars several hundred GeV in mass.  Since I have included in this scenario independent mass contributions for the scalars, the adjoint fermions and gauginos are free to be of order 100 GeV, while maintaining weak scale MSSM scalar masses.  Though scalar adjoints are heavy there remains the possibility of light adjoint fermions.  When some small R-breaking is introduced there is a mass splitting between a mostly-adjoint and mostly-gaugino states.  This may lead to the possibility of interesting cascade decays, and may shift current particle mass bounds.

\section{Appendix}
\begin{figure}[h]
\centerline{\includegraphics[width=7 cm]{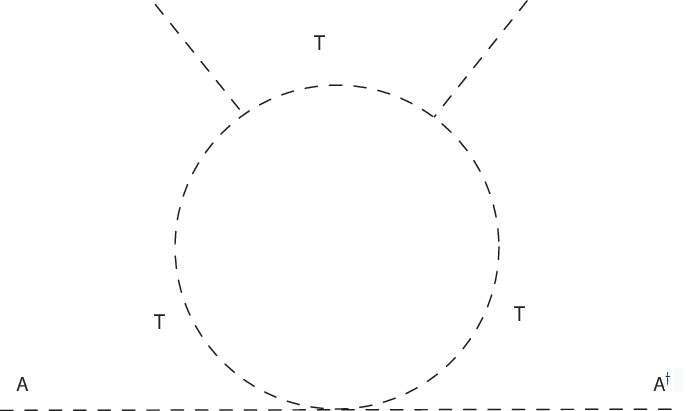}}
\label{fig:2pgm}
\end{figure}
\begin{figure}[h]
\centerline{\includegraphics[width=7 cm]{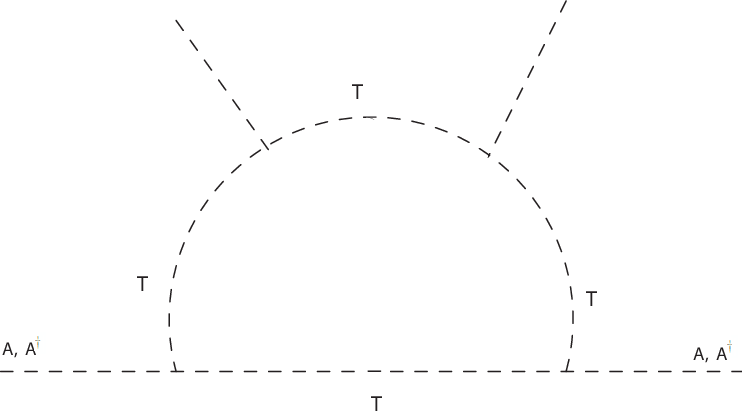}}
\label{fig:2pgm}
\end{figure}
One loop diagrams contributing to Adjoint soft masses. The first diagram is a mass $A A^{\dagger}$ while the second is a mass term
$AA+AA^{\dagger}+ h.c.$

\vskip1.5in

{\bf Acknowledgments}

This work was supported in part by NSF Grant No. PHY-0653656. Thanks to Yael Shadmi and Yuri Shirman and Ann Nelson for helpful discussions.

\end{document}